\documentclass[showpacs,preprintnumbers,amsmath,amssymb,prb]{revtex4}

\usepackage{graphicx}
\usepackage{dcolumn}
\usepackage{bm}

\begin{document}

\title{Unusual dynamics of spin-$\frac12$ antiferromagnets on the triangular lattice in magnetic field}

\author{A.\ V.\ Syromyatnikov}
\email{asyromyatnikov@yandex.ru}
\affiliation{Petersburg Nuclear Physics Institute named by B.P.\ Konstantinov of National Research Center "Kurchatov Institute", Gatchina 188300, Russia}

\date{\today}

\begin{abstract}

We theoretically discuss dynamical properties of spin-$\frac12$ Heisenberg antiferromagnet on the triangular lattice in magnetic field $\bf H$. We use the recently proposed bond-operator theory which operates with quantum states of the whole magnetic unit cell containing three spins. This technique describes accurately short-range spin correlations and provides a quantitative description of elementary excitations which appear in other approaches as bound states of conventional low-energy quasiparticles (e.g., magnons). In quantitative agreement with previous numerical and analytical findings, we observe four phases with coplanar spin arrangements upon the field increasing: the three-sublattice Y-phase, the collinear "up-up-down" (UUD) state, the non-collinear V-phase, and the collinear fully polarized (FP) state. We demonstrate that apart from magnons (spin-1 quasiparticles) there are spin-0 elementary excitations in the UUD state one of which is long lived and its spectrum lies below magnon branches. This mode originates from a high-energy quasiparticle at $H=0$ and it produces anomalies only in the longitudinal spin correlator because longitudinal and transverse channels are separated in collinear states. All other spin-0 excitations have finite lifetime and produce visibly anomalies in the dynamical structure factor. In the V-phase, we obtain multiple short-wavelength spin excitations which have no counterparts in the semiclassical spin-wave theory. Besides, we demonstrate a highly nontrivial field evolution of quasiparticles spectra on the way from one collinear state (UUD) to another one (FP) via the non-collinear V-phase (in which the longitudinal and the transverse channels are mixed). In particular, some parts of the spin-0 branch in the UUD state become parts of the spin-1 (magnon) branch in the FP phase whereas some parts of one magnon branch turn into parts of spin-2 branch. Such evolution would be very difficult to find by any conventional analytical approach. Our results are in good agreement with neutron experimental data obtained recently in $\rm Ba_3CoSb_2O_9$, $\rm KYbSe_2$, and $\rm CsYbSe_2$.

\end{abstract}


\maketitle

\section{Introduction}

Frustrated quantum magnetism has been one of the rapidly developing fields in condensed matter physics over the past decades because it offers a convenient playground for discussion of novel types of many-body phenomena including quantum spin-liquid phases, novel universality classes of phase transitions, and anomalous spin dynamics, to mention just a few. \cite{Balents2010,lacroix} Frustration arises due to lattice geometry when local spin interaction energies cannot be simultaneously minimized that leads to a large ground-state degeneracy. Quantum or thermal fluctuations can lift this degeneracy thereby selecting and stabilizing an ordered state (the order-by-disorder phenomenon). \cite{vill,shender2,henley1} Generally, coplanar and even collinear ground states are favored by this mechanism. 

The frustration can enrich the behavior of a system in the magnetic field $\bf H$ resulting in a number of unusual phenomena. In particular, quantum fluctuations at finite $H$ can stabilize collinear phases with gapped spectrum and magnetization plateaus, where  the magnetization does not change in a field interval below its saturation value $H_s$. \cite{lacroix}

\begin{figure}
\includegraphics[scale=1.]{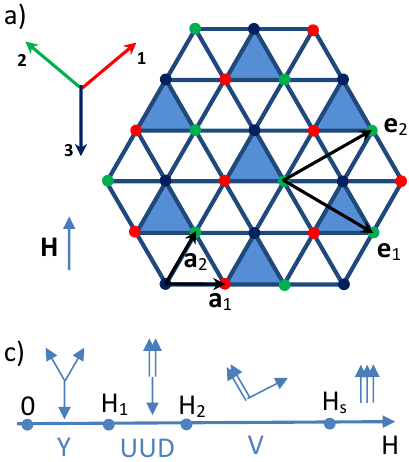}
\hskip 1cm
\includegraphics[scale=1.]{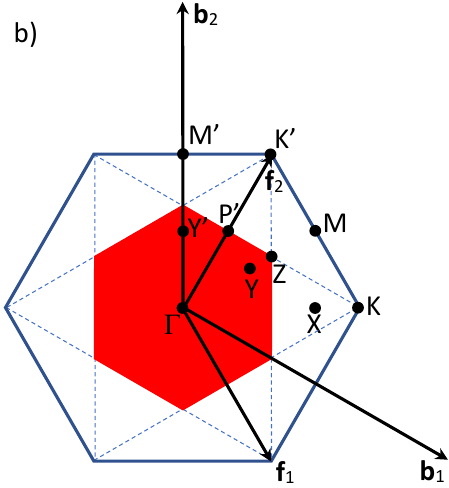}
\caption{
(a) Heisenberg spin-$\frac12$ antiferromagnet on the triangular lattice in magnetic field $\bf H$. Sites are distinguished by color belonging to three magnetic sublattices in the magnetically ordered phases. Translation vectors are shown of the crystal (${\bf a}_{1,2}$) and of the magnetic lattice (${\bf e}_{1,2}$).
(b) Crystal (blue hexagon) and magnetic (red hexagon) Brillouin zones. Translation vectors are depicted of the crystal (${\bf b}_{1,2}$) and of the magnetic (${\bf f}_{1,2}$) reciprocal lattices. Some high-symmetry points are marked.
(c) Phases arising in this model: "Y", "up-up-down" (UUD), "V", and the fully polarized phases. Orientations of three magnetic sublattices are shown in each state.
\label{BZfig}}
\end{figure}

Spin-$\frac12$ triangular-lattice Heisenberg antiferromagnet (HAF) has attracted a lot of interest as a typical frustrated system hosting many of the above mentioned phenomena. Its Hamiltonian has the form
\begin{equation}
\label{ham}
{\cal H} = J \sum_{\langle i,j \rangle}	{\bf S}_i{\bf S}_j  - H\sum_i S_i^z,
\end{equation} 
where $\langle i,j \rangle$ denote nearest-neighbor sites and the exchange coupling constant $J$ is set to be equal to unity below. Most theoretical studies have indicated that this model has an ordered ground state at $H=0$ with three magnetic sublattices shown in Fig.~\ref{BZfig}(a). \cite{egs1,egs2,egs3} At finite $H$, quantum and thermal fluctuations select states with coplanar spin arrangements presented in Fig.~\ref{BZfig}(c) from the manifold of non-collinear states having the same classical energy. \cite{chub91,kawa85} In particular, quantum fluctuations stabilize at $H_1<H<H_2$ the 1/3 magnetization plateau in the collinear up-up-down (UUD) state in which two sublattices align parallel to the field and the third sublattice is antiparallel to it. \cite{chub91} This theoretical prediction was confirmed numerically, \cite{honecker,farnell,sakai,starykh,danshita,sakai2} analytically \cite{plattri}, and experimentally in spin-$\frac12$ materials $\rm Cs_2CuBr_4$ \cite{cscubr} and $\rm Ba_3CoSb_2O_9$ \cite{Doi_2004,bacoH,bacogap,bacoprl}, and in spin-$\frac52$ compound $\rm RbFe(MoO_4)_2$ \cite{inami96}.

Despite quite definite success in search and characterization of phases induced by fluctuations in frustrated systems, much less is known about their dynamical properties. This point is of particular interest for the following reasons. There is a growing number of experimental and numerical evidences that conventional theoretical approaches do not describe even qualitatively short-wavelength elementary excitations in ordered phases of both frustrated and non-frustrated spin-$\frac12$ systems. For instance, it was found numerically in spin-$\frac12$ HAF on the square lattice in strong magnetic field below its saturation value $H_s$ that a number of peaks appear in the dynamical structure factor (DSF) at a given momentum instead of one magnon peak predicted by the spin-wave theory (SWT). \cite{olav,magfail2} Then, in recent inelastic neutron scattering experiments performed in $\rm Ba_3CoSb_2O_9$, the inability was demonstrated of standard theoretical approaches to describe short-wavelength spin excitations in model \eqref{ham} at $H=0$. \cite{triang1,bacoprl,bacoprb} In particular, neutron scattering data show at least four peaks at $M$ point (see Fig.~\ref{BZfig}(b)) of the Brillouin zone (BZ), while the SWT predicts only two magnon peaks and a high-energy continuum of excitations. \cite{chub_triang,zh_triang,zhito}

To resolve these problems in the theory, we have proposed recently a new method based on the bond representation of spin-$\frac12$ operators in terms of Bose operators. \cite{ibot,aktersky,iboth,itri} This bond operator theory (BOT) which is described in some detail in Sec.~\ref{secbot} is very close in spirit to the standard SWT but it more accurately takes into account short-range spin correlations and makes it possible, along with magnons, to quite simply study high-energy excitations arising in the SWT as bound states of several magnons. In spin-$\frac12$ HAF on the square lattice in strong field, we demonstrate by the BOT that quantum fluctuations are so strong in this system that the numerous anomalies in dynamical spin correlators correspond to poles of Green's functions which have no counterparts in the semiclassical SWT. \cite{iboth} In triangular-lattice HAF \eqref{ham} in zero field, we show that quantum fluctuations considerably change properties of conventional magnons predicted by the SWT. \cite{itri} Besides, we observe novel high-energy collective excitations built from high-energy excitations of the magnetic unit cell (containing three spins) and another novel high-energy quasiparticle which has no counterpart not only in the SWT but also in the harmonic approximation of the BOT. All observed elementary excitations produce visible anomalies in dynamical spin correlators and describe experimental data obtained in $\rm Ba_3CoSb_2O_9$. \cite{itri}

Thus, it is interesting to consider by the BOT the problem of elementary excitations in spin-$\frac12$ triangular-lattice HAF in finite field \eqref{ham} that is the scope of the present paper. We discuss static properties of model \eqref{ham} in Sec.~\ref{static} and show that critical fields $H_{1,2}$ (see Fig.~\ref{BZfig}(c)), the ground-state energy, and the uniform magnetization are in a quantitative agreement with previous numerical findings.

We consider the spin dynamics in Sec.~\ref{dyn}. It is shown that one of the high-energy excitation, which contribute to the high-energy anomaly in the DSF at $H=0$, moves down upon the field increasing and turns into a spin-0 quasiparticle in the UUD state whose spectrum lies below all magnon modes. We find also other spin-0 excitations which have finite lifetime and some of which have no counterpart in the harmonic approximation of the BOT. All spin-0 modes produce anomalies in the longitudinal two-spin correlator whose spectral weights, however, are much smaller than spectral weights of magnons (spin-1 excitations) in the transverse spin correlators that hinders their experimental observation. We demonstrate the appearance of multiple short-wavelength spin excitations at $H_2<H<H_s$ (i.e., in the V-phase shown in Fig.~\ref{BZfig}(c)) producing sharp anomalies in two-spin dynamical correlators which have no counterparts in the SWT. We find also a highly nontrivial spectra evolution upon the field increasing. In particular, some parts of the low-lying spin-0 branch in the UUD state become parts of the spin-1 (magnon) branch in the fully polarized phase after passing through the non-collinear V-phase (in which the longitudinal and the transverse channels are mixed). Then,  some parts of one magnon branch turn into parts of spin-2 branch. Such evolution would be very difficult to find by any conventional analytical approach whereas the BOT allows to detect it even in the harmonic approximation.

Sec.~\ref{expersec} is devoted to comparison of our theory predictions with available experimental and numerical findings. We add to model \eqref{ham} a small easy-plane anisotropy, calculate dynamical two-spin correlators, and compare our findings with available inelastic neutron scattering data observed in the UUD phase of $\rm Ba_3CoSb_2O_9$ and with results of SWT calculations reported in Ref.~\cite{bacoH}. We obtain that magnons produce pronounced anomalies in the neutron cross section in good agreement with the experiment and with the SWT. We demonstrate that the spectral weight of the low-lying spin-0 excitations is too small to be detected in the experiment with non-polarized neutrons carried out in Ref.~\cite{bacoH}.

We also show in Sec.~\ref{expersec} that our results are in good agreement with experimental and numerical data obtained in KYbSe$_2$ \cite{kybse} and CsYbSe$_2$ \cite{csybse1,csybse2}. In the UUD state, energies of all magnons (transverse excitations) found within the BOT are in good quantitative agreement with the experiment and with numerical calculations performed in Ref.~\cite{csybse2} using matrix-product-state (MPS) representations. The agreement in the longitudinal channel is also good and it allows us to propose that longitudinal excitations which we obtain within the BOT in the UUD phase are really observed experimentally and numerically in Ref.~\cite{csybse2}. In the V state, the appearance within the BOT of separately standing anomalies in the DSF which are produced by new quasiparticles is also consistent with numerical results.

Sec.~\ref{conc} contains a summary of results and our conclusions.

\section{Bond-operator formalism for triangular-lattice spin-$\frac12$ magnets and general consideration}
\label{secbot}

\begin{figure}
\includegraphics[scale=0.4]{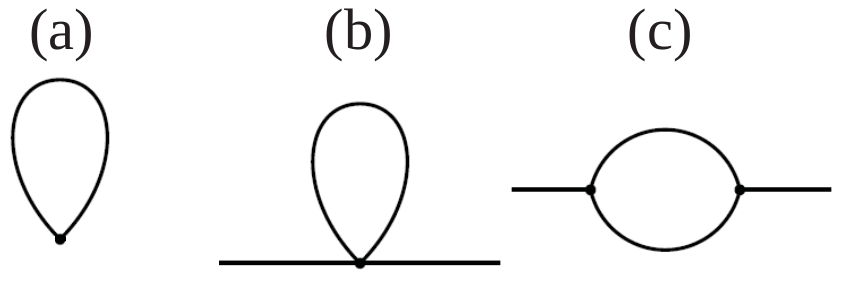}
\caption{(a) The diagram giving corrections of the first-order in $1/n$ to the ground state energy and to the staggered and uniform magnetizations. (b), (c) Diagrams of the first order in $1/n$ for self-energy parts.
\label{diag}}
\end{figure}

The general procedure is described in detail in Ref.~\cite{ibot} for building of the bosonic spin representation for more than one spin 1/2 in the unit cell. We use three-spin variant of this representation in the current study for discussion of all ordered phases arising in model \eqref{ham} (see Fig.~\ref{BZfig}) which have three spins in the magnetic unit cell. We introduce seven Bose operators in each magnetic unit cell acting on eight basis functions $|0\rangle$ and $|e_i\rangle$ ($i=1,...,7$) of three spins according to the rule
\begin{equation}
\label{bosons}
	a_i^\dagger |0\rangle = |e_i\rangle, \quad i=1,...,7,
\end{equation}
where $|0\rangle$ is a selected state playing the role of the vacuum. Suitable basis functions are presented in Ref.~\cite{itri} devoted to the considered model at $H=0$.
\footnote{
The only difference with Ref.~\cite{itri} is that we have to introduce six parameters $\alpha$ controlling the mixing of the basis functions in order to describe all phases arising at $H\ne0$. See Refs.~\cite{ibot,itri} for a more detailed discussion.
} 
Then, we build the bosonic representation of spins in the unit cell as it is described in Refs.~\cite{ibot,itri} which is too bulky to be presented here. There is a formal artificial positive parameter $n$ in this representation arising in operator $\sqrt{n-\sum_{i=1}^7a_i^\dagger a_i}$ by which linear in Bose operators terms are multiplied (cf.\ the term $\sqrt{2S-a^\dagger a}$ in the Holstein-Primakoff representation). Due to this operator, Bose analogs of spin operators do not mix states containing more than $n$ bosons in the unit cell and states with no more than $n$ bosons. Then, the physical results of the BOT correspond to $n=1$. Bilinear in Bose operators terms do not depend on $n$ and have the form $a_i^\dagger a_j$ in our spin representation and all constant terms are proportional to $n$. We introduce also separate representations via operators \eqref{bosons} for terms ${\bf S}_i{\bf S}_j$ in the Hamiltonian in which $i$ and $j$ belong to the same unit cell. Constant terms in these representations are proportional to $n^2$ and terms of the form $a_i^\dagger a_j$ are proportional to $n$. \cite{ibot} As a result, we come to a close analog of the Holstein-Primakoff spin transformation in which $n$ is the counterpart of the spin value $S$ and which reproduces the spin commutation algebra for all $n>0$. Similar to the SWT, expressions for observable quantities can be found within the BOT as series in $1/n$ using the conventional diagrammatic technique because terms in the Bose-analog of the spin Hamiltonian containing products of $q$ Bose operators are proportional to $n^{2-q/2}$ (such terms are proportional to $S^{2-q/2}$ in the SWT). In particular, one has to consider diagrams presented in Fig.~\ref{diag} to calculate the ground-state energy, the staggered and the uniform magnetizations, and self-energy parts in the first order in $1/n$ (as it has to be done in the first order in $1/S$ in the SWT). 

It is clear that apart from bosons describing well-known quasiparticles (magnons in model \eqref{ham}) there are extra Bose operators in the BOT which describe spin excitations appearing in conventional approaches as bound states of the common quasiparticles. \cite{ibot,inem} Interestingly, spectra of all bosons are obtained in the BOT as series in $1/n$ by calculating the same diagrams. In contrast, an analysis is required of an infinite series of diagrams for multi-particle vertexes in common methods to find spectra of magnon bound states. Our previous applications of the BOT to some two-dimensional models well studied before by other numerical and analytical methods show that in most cases first $1/n$ terms give the main contributions to renormalization of all observables if the system is not too close to a quantum critical point (similarly, first $1/S$ corrections in the SWT frequently make the main quantum renormalization of observable quantities far from quantum critical points even in 2D spin-$\frac12$ systems, Ref.~\cite{monous}). \cite{ibot,aktersky,itri,iboth} In particular, the description by the BOT of all short-wavelength quasiparticles appears to be more precise compared to the SWT indicating that short-range spin correlations are taken into account in the BOT more accurately. Importantly, because the spin commutation algebra is reproduced in the BOT at any $n>0$, the proper number of Goldstone excitations arises in ordered phases in any order in $1/n$ (unlike the vast majority of other versions of BOT proposed so far, see Ref.~\cite{ibot}). The main disadvantage of the BOT is that it is very bulky (e.g., there are more than 100 terms in the part of Hamiltonian \eqref{ham} bilinear in Bose operators). As a result, the numerical calculation of diagrams takes a lot of time. 
%That is why there is a limited number of points on some plots below found in the first order in $1/n$.

Some general statements can be made about observable quantities to be found within the BOT in model \eqref{ham}. Both in the SWT and in the BOT, there should be Goldstone excitations at $\Gamma$ and $K$ points of the BZ (see Fig.~\ref{BZfig}(b)) in phases with broken continuous symmetry in all orders in $1/S$ and $1/n$. In the UUD and in the fully polarized phases, the rotational symmetry along the field direction is unbroken and all excitations should be gapped. \cite{chub91} As the projection of the total spin $S^z$ on $z$ axis commutes with Hamiltonian \eqref{ham}, all system states can be characterized by their $S^z$ value in the collinear phases. That is why the uniform magnetization does not depend on $H$ (showing plateaus) in the collinear states and we can distinguish excitations by the $S^z$-sector in which they live (e.g., spin-0 excitations, spin-1 excitations, etc.). In contrast, $S^z$ is not a good quantum number in non-collinear $Y$ and $V$ phases.

Then, there should exist a homogeneous mode at $\Gamma$ and $K$ points with zero damping whose energy is equal to $H$. \cite{golosov,chub91} As the energy of this mode can be derived using only the spin commutation relations (see Ref.~\cite{golosov}), its energy should be equal to $H$ in all orders in $1/S$ and $1/n$ because the spin commutation algebra is fulfilled at any $S$ and $n$ in the SWT and in the BOT, respectively. This mode is related to the precession of the uniform magnetization of the system around the field.

Another general statement can be made about the fully polarized phase. There are no zero-point fluctuations at $H\ge H_s$ both in the SWT and in the BOT so that there are no $1/S$ and $1/n$ corrections to static observable quantities and to magnon spectra: all diagrams are zero at $T=0$ because they contain closed contours which can be walked around along arrows of Green's functions and which give zero upon the integration over frequencies. Then, results are precise for static observables and magnon spectra obtained in the fully polarized phase in the harmonic approximations of the SWT and of the BOT. However, there are three-particle vertexes in the BOT at $H\ge H_s$ describing the decay of high-energy excitations carrying spin 2 into two spin-1 quasiparticles (magnons). Similarly, there are three-particle vertexes describing the decay of high-energy spin-3 elementary excitation into spin-1 and spin-2 quasiparticles. Then, there are finite $1/n$ corrections to spectra of spin-2 and spin-3 elementary excitations at $H\ge H_s$. These spin-2 and spin-3 quasiparticles correspond in the SWT to bound states of two and three magnons, respectively. Their spectra are given in the SWT by poles of four- and six-particle vertexes whereas they are described by separate bosons in the BOT whose spectra can be found in the same way as it is done for magnons (i.e., by considering $1/n$ corrections to their spectra from diagrams shown in Figs.~\ref{diag}(b) and \ref{diag}(c)). Spin-2 excitations will be discussed in more detail in Sec.~\ref{sec1n}.

\section{Static properties}
\label{static}

In agreement with previous analytical and numerical findings, we obtain using the BOT four phases in the considered model shown in Fig.~\ref{BZfig}(c): two non-collinear Y and V states with broken continuous rotational symmetry in the plane perpendicular to $\bf H$ and two collinear phases (the UUD and the fully polarized ones) in which this symmetry is unbroken. Transitions between these phases are of the second order: transformations of spin orderings are smooth in the first two orders in $1/n$ and one of the excitation branches becomes soft upon transitions from gapped UUD and the fully polarized phases to Y and V states as it is demonstrated below. 

\begin{figure}
\includegraphics[scale=1.1]{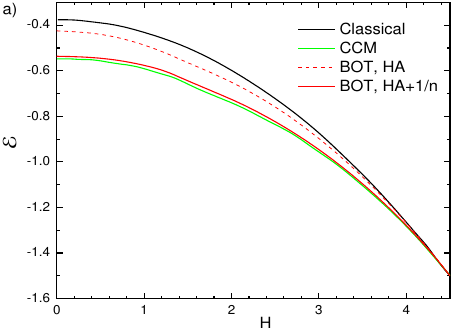}
\includegraphics[scale=1.1]{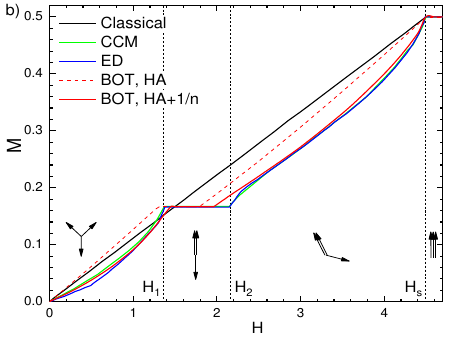}
\caption{
The ground state energy ${\cal E}$ a) and the uniform magnetization $M$ per spin b) obtained in the linear spin-wave theory (classical), within the high-order coupled cluster method (CCM) \cite{farnell}, using the exact diagonalization of finite clusters with the subsequent continuation to the thermodynamic limit (ED) \cite{honecker}, and within the BOT in the harmonic approximation (HA) and in the first order in $1/n$ (HA+$1/n$). $H_{1,2}$ are given by Eqs.~\eqref{h12} in the BOT. Orientation of three sublattices are shown in insets in panel b).
\label{emfig}}
\end{figure}

Plots are shown in Fig.~\ref{emfig} of the ground state energy ${\cal E}$ and the uniform magnetization $M$ per spin found in the harmonic approximation of the BOT and in the first order in $1/n$ as it is explained in detail in Refs.~\cite{ibot,itri}. Corresponding results are also presented in Fig.~\ref{emfig} obtained in the linear SWT and using numerical methods.

It is seen from Fig.~\ref{emfig}(a) that quantum fluctuations lower the ground state energy. Notice that in contrast to the SWT some amount of quantum fluctuations are taken into account even in the harmonic approximation of the BOT. We stress also that first corrections in $1/n$ bring our results for $\cal E$ to the quantitative agreement with previous numerical findings making the main renormalization of this quantity in the BOT (as it happens in all other systems considered before by the BOT which are not too close to a quantum critical point \cite{ibot,aktersky,itri}).

In contrast to the linear spin-wave theory, the minimization of the bare ground-state energy (given by the term in the Hamiltonian not containing Bose operators) does give the UUD phase (with the 1/3-magnetization plateau) in the BOT in the finite field interval $H_1\le H\le H_2$ (see Fig.~\ref{emfig}(b)). This is not very surprising because some amount of quantum fluctuations is taken into account in the harmonic approximation of the BOT while it is known that the UUD phase is stabilized by quantum or thermal fluctuations. Calculation of transverse spin components gives for critical fields in the first order in $1/n$
\begin{equation}
\label{h12}
\begin{array}{ll}
	&H_1 = 1.3n + 0.15, \\
	&H_2 = 1.8n + 0.1.
\end{array}
\end{equation}
Similar to the ground state energy, corrections of the first order in $1/n$ make the main renormalization of the uniform magnetization providing the quantitative agreement with numerical results (see Fig.~\ref{emfig}(b)). Some discrepancy is seen only near the critical fields $H_{1,2}$ signifying a slower convergence of $1/n$ series near these quantum critical points.

\section{Dynamical properties}
\label{dyn}

We calculate in this section dynamical spin susceptibilities
\begin{eqnarray}
\label{chi}
\chi_{\alpha\beta}({\bf k},\omega) &=&
i\int_0^\infty dt 
e^{i\omega t}	
\left\langle \left[ S^\alpha_{\bf k}(t), S^\beta_{-\bf k}(0) \right] \right\rangle,\\
\label{chizz}
\chi_\|({\bf k},\omega) &=& \chi_{zz}({\bf k},\omega),\\
\label{chiperp}
\chi_\perp({\bf k},\omega) &=& \chi_{xx}({\bf k},\omega) +  \chi_{yy}({\bf k},\omega),\\
\label{chitot}
\chi_{tot}({\bf k},\omega) &=& \chi_{xx}({\bf k},\omega) +  \chi_{yy}({\bf k},\omega) + \chi_{zz}({\bf k},\omega),
\end{eqnarray}
and dynamical structure factors (DSFs)
\begin{eqnarray}
\label{dsfzz}
{\cal S}_\|({\bf k},\omega) &=& 
\frac1\pi {\rm Im}
\chi_\|({\bf k},\omega),\\
\label{dsfperp}
{\cal S}_\perp({\bf k},\omega) &=& 
\frac1\pi {\rm Im}
\chi_\perp({\bf k},\omega),\\
\label{dsftot}
{\cal S}_{tot}({\bf k},\omega) &=& 
\frac1\pi {\rm Im}
\chi_{tot}({\bf k},\omega),
\end{eqnarray}
where
\begin{equation}
\label{sk}
	{\bf S}_{\bf k} = \frac{1}{\sqrt3} 
	\left( 
	{\bf S}_{1\bf k} + {\bf S}_{2\bf k}e^{-i(k_1+k_2)/3} + {\bf S}_{3\bf k}e^{-i(2k_2-k_1)/3}
	\right)
\end{equation}
are built on spin operators 1, 2, and 3 in the unit cell (see Fig.~\ref{BZfig}(a)), 
${\bf k} = k_1{\bf f}_1 + k_2{\bf f}_2$, and ${\bf f}_{1,2}$ are depicted in Fig.~\ref{BZfig}(b). Poles of $\chi_{\alpha\beta}({\bf k},\omega)$ determine spectra of spin excitations in the system and produce anomalies in DSFs. 

\begin{figure}
\includegraphics[scale=0.08]{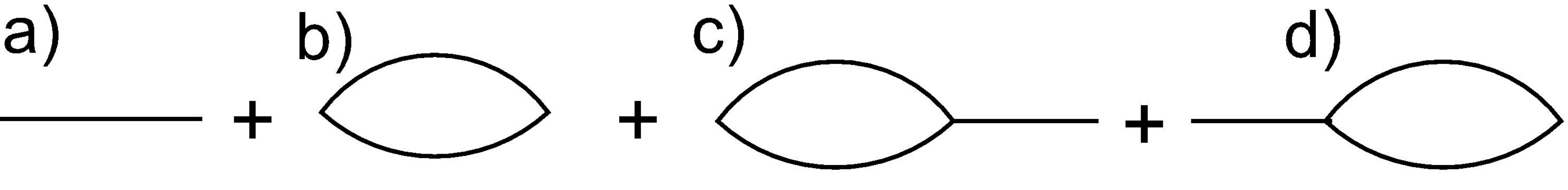}
\caption{Diagrams for spin susceptibilities \eqref{chi} to be taken into account in the first order in $1/n$.
\label{chifig}}
\end{figure}

In the leading order in $1/n$ (i.e., in the harmonic approximation), $\chi_{\alpha\beta}({\bf k},\omega)$ have the form of linear combinations of Green's functions of bosons. Strictly speaking, one has to take into account diagrams shown in Fig.~\ref{chifig} in the consideration of spin susceptibilities in the first order in $1/n$. However, we take into account only the diagram shown in Fig.~\ref{chifig}(a) below and consider $1/n$ corrections to self-energy parts. The main reason for this restriction is that our main aim is the consideration of quasiparticles spectra whereas diagrams shown in Figs.~\ref{chifig}(b)--\ref{chifig}(d) either renormalize quasiparticles spectral weights or contribute to the incoherent background. Besides, all calculations in the first order in $1/n$ are quite time-consuming. The diagram shown in Fig.~\ref{chifig}(b) is taken into account only in Sec.~\ref{expersec} for better reproducing the incoherent background observed experimentally and numerically.

\subsection{Harmonic approximation}

\begin{figure}
\includegraphics[scale=0.88]{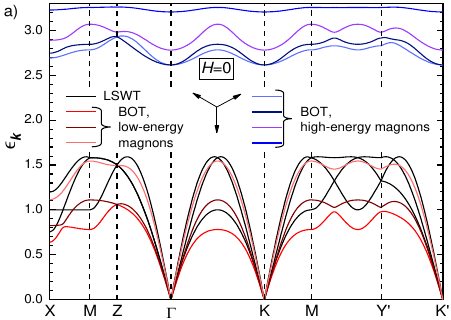}
\includegraphics[scale=0.88]{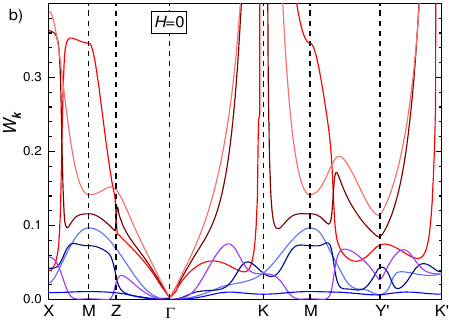}
\includegraphics[scale=0.88]{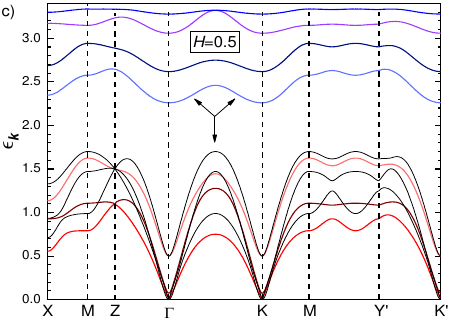}
\includegraphics[scale=0.88]{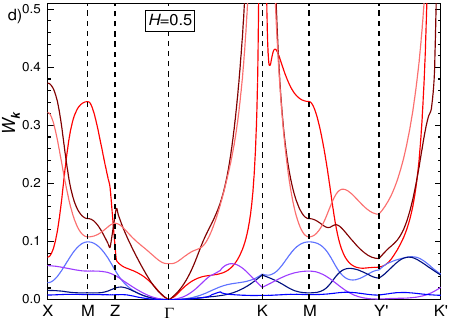}
\includegraphics[scale=0.88]{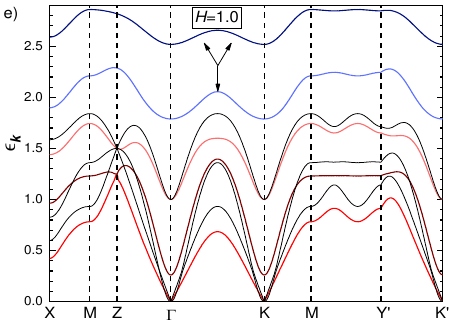}
\includegraphics[scale=0.88]{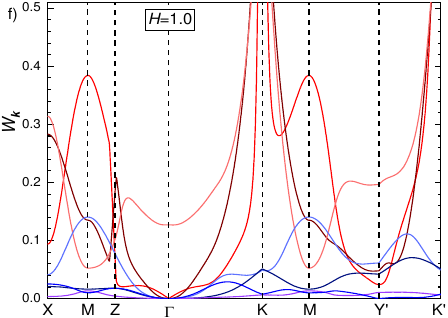}
\includegraphics[scale=0.88]{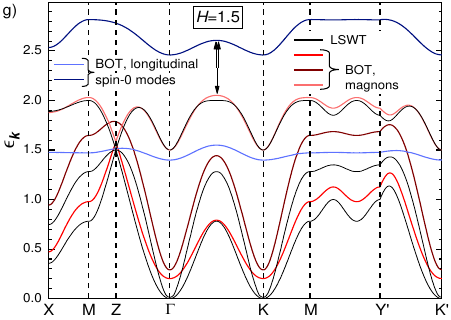}
\includegraphics[scale=0.88]{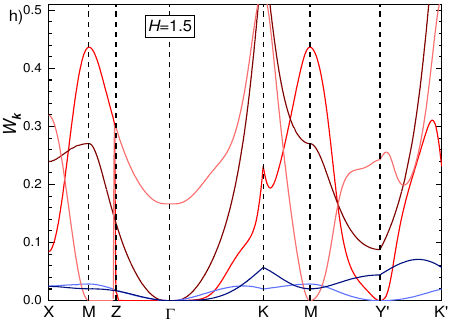}
\includegraphics[scale=0.88]{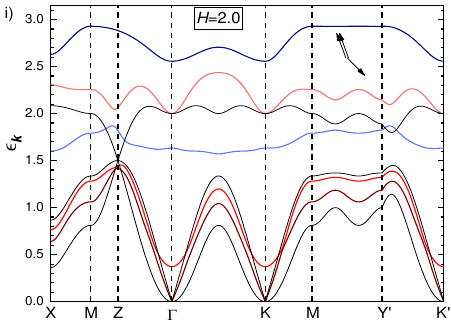}
\includegraphics[scale=0.88]{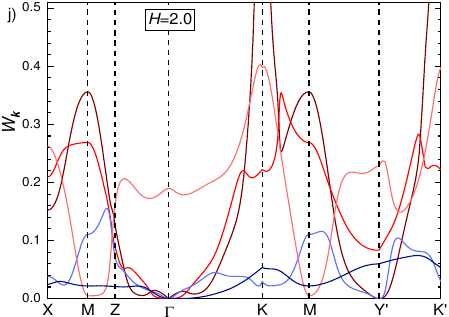}
\caption{
Spectra $\epsilon_{\bf k}$ of low-lying excitations found in the linear spin-wave theory (LSWT) and in the harmonic approximation of the BOT in the field interval $0\le H\le2$. Spectral weights $W_{\bf k}$ are also presented of all quasiparticles in dynamical structure factor \eqref{dsftot} found within the harmonic approximation of the BOT. The path along the Brillouin zone goes through high-symmetry points marked in Fig.~\ref{BZfig}(b). Orientation of three magnetic sublattices are depicted in insets for each $H$ value.
\label{spec01}}
\end{figure}

\begin{figure}
\includegraphics[scale=0.9]{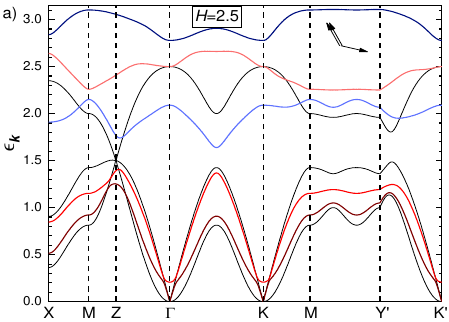}
\includegraphics[scale=0.9]{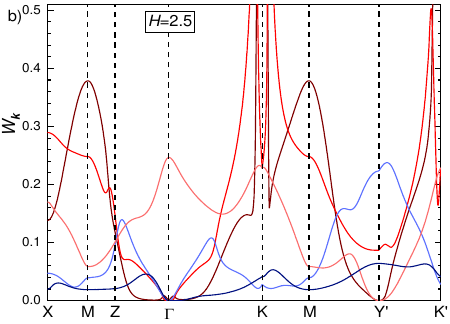}
\includegraphics[scale=0.9]{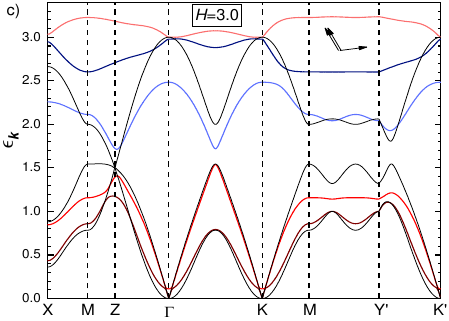}
\includegraphics[scale=0.9]{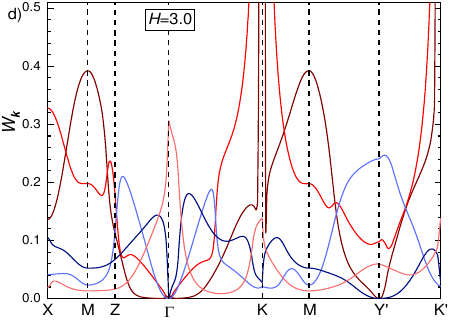}
\includegraphics[scale=0.9]{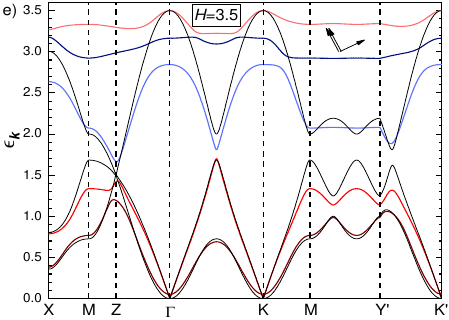}
\includegraphics[scale=0.9]{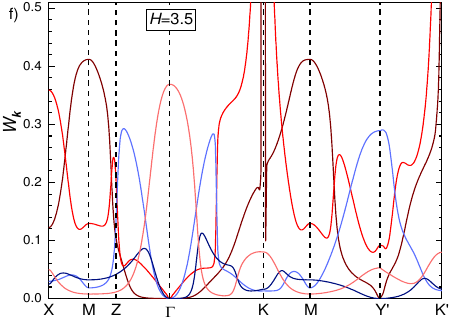}
\includegraphics[scale=0.9]{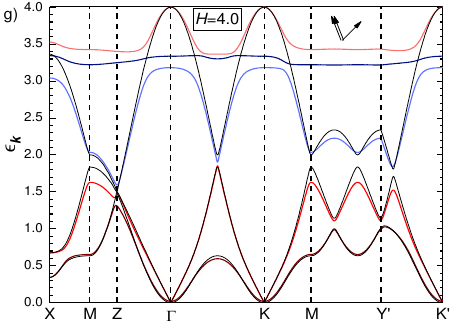}
\includegraphics[scale=0.9]{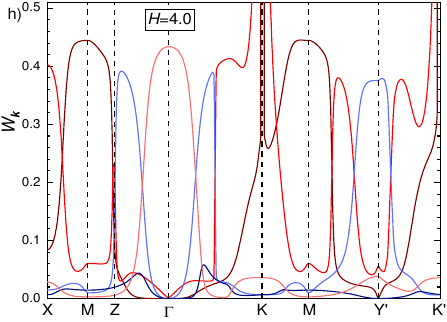}
\includegraphics[scale=0.9]{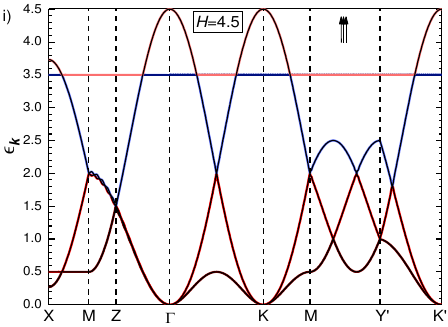}
\includegraphics[scale=0.9]{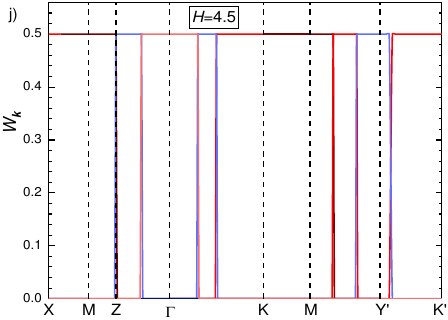}
\caption{
Same as Fig.~\ref{spec01} but for the field interval $2.5\le H\le H_s=9/2$.
\label{spec02}}
\end{figure}

Spectra of low-lying elementary excitations found in the harmonic approximation of the BOT are shown in Figs.~\ref{spec01} and \ref{spec02} in field intervals $0\le H\le2$ and $2.5\le H\le H_s=9/2$, respectively. Spectral weights $W_{\bf k}$ of all quasiparticles in ${\cal S}_{tot}({\bf k},\omega)$ (i.e., coefficients before corresponding delta-functions in Eq.~\eqref{dsftot}) are also presented in Figs.~\ref{spec01} and \ref{spec02}. For convenience of comparison, we present in Figs.~\ref{spec01} and \ref{spec02} also magnon spectra found in the linear SWT in a standard way. There are three magnon branches in the SWT corresponding to three spins in the magnetic unit cell.

\underline{$H=0$.} It is shown in our previous paper Ref.~\cite{itri} that all seven quasiparticles arising in the harmonic approximation of the BOT produce visible anomalies in ${\cal S}_{tot}({\bf k},\omega)$ (see Fig.~\ref{spec01}(b)). Then, we call all of them magnons in Ref.~\cite{itri}: there are three low-energy Goldstone magnons corresponding to conventional spin waves in the SWT and there are four high-energy magnons (see Fig.~\ref{spec01}(a)). In contrast to the SWT in which magnons are elementary excitations related with spins fluctuations transverse to the sublattices magnetizations, all quasiparticles are of mixed type (i.e., both longitudinal and transverse) in the BOT at $H=0$ except for the highest-energy low-energy magnon and the highest-energy high-energy magnon which are purely transverse. As it is discussed in detail in Ref.~\cite{itri}, quantum fluctuations lift the degeneracy of low-energy magnon branches predicted by the SWT \cite{chub_triang,zh_triang,zhito} along $\Gamma M$ lines and along blue dashed lines depicted in Fig.~\ref{BZfig}(b) (notice that this degeneracy remains in the SWT even in the first order in $1/S$). This our conclusion is in agreement with recent experiments in $\rm Ba_3CoSb_2O_9$ at $H=0$. \cite{itri}

\underline{$0<H<H_1$.} It is seen from Figs.~\ref{spec01}(a), \ref{spec01}(c), \ref{spec01}(e), and \ref{spec01}(g) that four high-energy branches of excitations behave differently in the Y-phase upon the field increasing: energies of two of them increase (we do not show them in Fig.~\ref{spec01} at $H\ge1$), one high-energy branch moves down, and the energy of the remaining branch does not practically change. Two of low-energy magnons in the BOT and two magnon branches in the SWT acquire gaps at $\Gamma$ and $K$ points one of which is equal to $H$ (it is discussed above). Only one Goldstone magnon remains in the Y-phase.

\underline{$H_1<H<H_2$.} Spectra are shown in Fig.~\ref{spec01}(g) of low-lying quasiparticles obtained in the BOT in the UUD phase. Three low-energy magnons become purely transverse (i.e., they produce anomalies only in DSF \eqref{dsfperp}) in this phase and the remaining two low-lying elementary excitations whose spectra are shown in Fig.~\ref{spec01}(g) in light blue and blue lines are longitudinal (i.e., they produce anomalies only in DSF \eqref{dsfzz}) and they carry spin 0. As far as we know, the latter two longitudinal spin-0 excitations have not been discussed yet in the UUD state. They could appear in the SWT in the two-particle channel as bound states of two magnons and would be related with poles in four-particle vertexes. Interestingly, the magenta line shown in Fig.~\ref{spec01}(a) and \ref{spec01}(c) turns into high-energy branch of spin-2 excitations in the UUD phase.

\begin{figure}
\includegraphics[scale=1.1]{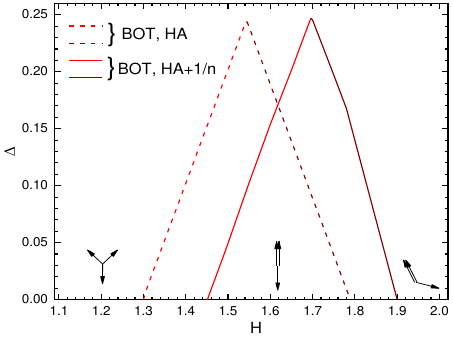}
\caption{
Energy gap to the first excited state found in the BOT within the harmonic approximation (HA) and in the first order in $1/n$ (HA$+1/n$). The gap is finite in the UUD collinear phase arising in the interval $H_1<H<H_2$, where the 1/3-plateau appears in the uniform magnetization $M$ (see Fig.~\ref{emfig}(b)). Lines colors correspond to branches shown in Fig.~\ref{spec01}.
\label{gapfig}}
\end{figure}

As there no continuous symmetry breaking in the UUD phase, there is a gap to the lowest excited state (at $\Gamma$ and $K$ points) which is plotted in Fig.~\ref{gapfig}. In agreement with conclusions made within the SWT \cite{chub91}, different low-energy branches become gapless in the BOT at $H=H_1$ and $H=H_2$. We point out also at least qualitative agreement at $0<H<H_2$ between magnon branches in the SWT and three low-energy branches (stemming from three low-energy magnons at $H=0$) in the BOT.

\underline{$H_2<H\le H_s$.} As it is seen from Figs.~\ref{spec01} and \ref{spec02}, the close similarity remains in the V-phase only between two lowest-energy branches obtained in the SWT and in the BOT (i.e., at $\omega<1.5$). In particular, there is only one Goldstone magnon both in the SWT and in the BOT. 

The field evolution of the magnon spectrum at $\omega>1.5$ in the SWT differs drastically from evolution of the branch in the BOT which stems from the highest-energy Goldstone magnon at $H=0$ (it is drawn by orange in Figs.~\ref{spec01} and \ref{spec02}). These branches coincide only at $\Gamma$ and $K$ points, where they are equal to $H$ as it is explained above but their overall behavior differs drastically at $2<H\le H_s$. In particular, it is seen from Fig.~\ref{spec02}(i) that at $H=H_s=9/2$, when results coincide for three magnon branches observed in the SWT and in the BOT (see above), the magnon branch is composed in BOT at $\omega>1.5$ from parts of two branches: from the band which was the highest-energy Goldstone magnon at $H=0$ and from the branch which was purely longitudinal in the UUD state (these branches are shown, respectively, in orange and in light blue in Figs.~\ref{spec01} and \ref{spec02}). Besides, horizontal parts of the "orange" and the "light blue" branches (at $\omega=3.5$) describe purely spin-2 excitations at $H=H_s$ (see Fig.~\ref{spec02}(i)) which can produce anomalies only in a four-spin correlator (see Fig.~\ref{spec02}(j) and discussion below). 

Thus, in terms of the SWT, we obtain using the BOT quite unexpected nontrivial interplay at $H<H_s$ between one-particle and two-particle sectors. However, this interplay cannot be reproduced by the SWT in first few orders in $1/S$ because consideration of the two-magnon channel requires within the SWT a discussion of some infinite series of diagrams for the four-particle vertex.

\subsection{First order in $1/n$}
\label{sec1n}

We calculate now DSFs and quasiparticles spectra taking into account corrections to the self-energy parts of the first order in $1/n$. It should be stressed that we take into account $\omega$-dependence of all self-energy parts and do not expand in $1/n$ neither numerator nor denominator of bosons Green's functions near bare poles. In our previous considerations \cite{itri,iboth} by the BOT of model \eqref{ham} at $H=0$ and of the square-lattice HAF in strong field, this allowed to observe the non-trivial renormalization of Green's functions denominator and to obtain, in particular, new poles corresponding to new quasiparticles in agreement with numerical and experimental findings. By varying $n$ value, we trace below the poles evolution from the harmonic approximation ($n\to\infty$) to physical results at $n=1$ that helps to identify novel poles which have no counterparts in the harmonic approximation.

\begin{figure}
\includegraphics[scale=1.2]{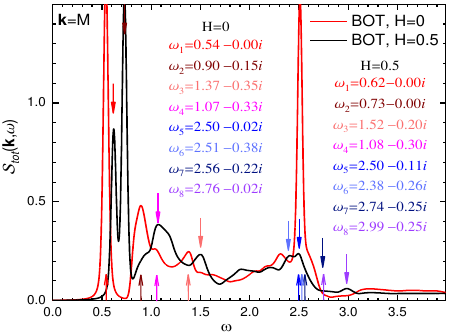}
\caption{
Dynamical structure factor (DSF) \eqref{dsftot} at point $M$ of the BZ (see Fig.~\ref{BZfig}(b)) at $H=0$ and $H=0.5$. DSF obtained within the first order in $1/n$ has been convoluted with the energy resolution $0.02J$. Anomalies in the DSF are produced by poles of spin correlator \eqref{chitot} indicated in insets by colors corresponding to excitation branches shown in Fig.~\ref{spec01}. Real parts of these poles are marked by upward and downward arrows of corresponding colors for $H=0$ and $H=0.5$, respectively. Pole $\omega_4$ has no counterpart neither in the spin-wave theory nor in the harmonic approximation of the BOT.
\label{dsfm005}}
\end{figure}

\underline{$H=0$.} The zero field limit was discussed in detail in our previous study Ref.~\cite{itri}. The most striking difference with conclusions of existing theories was obtained experimentally near $M$ point of the BZ (see Fig.~\ref{BZfig}(b)): at least four pronounced anomalies were observed experimentally whereas, for instance, the SWT predicts two magnon peaks and an incoherent continuum of excitations. Our calculations of DSFs in Ref.~\cite{itri} reproduced all anomalies found experimentally. ${\cal S}_{tot}({\bf k},\omega)$ obtained in Ref.~\cite{itri} at $M$ point at $H=0$ is shown in Fig.~\ref{dsfm005}. We observed that quantum fluctuations lift the magnon spectra degeneracy along $\Gamma M$ lines so that three low-energy peaks were produced by magnons (poles $\omega_{1,2,3}$ in Fig.~\ref{dsfm005}). The fourth low-energy anomaly (produced by pole $\omega_4$) appears at $n\approx2$ only after taking into account self-energy parts and it has no counterparts neither in the SWT nor in the harmonic approximation of the BOT. Pole $\omega_4$ has quite large imaginary part which, however, can be reduced to zero by small easy-plane anisotropy in the system. \cite{itri} The high-energy anomaly in DSFs at $\omega\approx2.5$ is produced at $M$ point by four high-energy magnons some of which are well defined quasiparticles (see Fig.~\ref{dsfm005}). This anomaly was really observed experimentally in $\rm Ba_3CoSb_2O_9$. \cite{triang1,bacoprl,bacoprb}

\underline{$0<H<H_1$.} We show in Fig.~\ref{dsfm005} also ${\cal S}_{tot}({\bf k},\omega)$ at $M$ point at $H=0.5$ to illustrate peaks evolution at small fields. The most pronounced difference with the case of zero field is that magnon pole $\omega_2$ moves to smaller energies leaving the two-magnon continuum so that its damping becomes zero. The high-energy anomaly becomes broader because energies of high-energy magnons disperse.

We point out also that spectra of short-wavelength (high-energy) excitations do not obey the sixfold rotation symmetry of the BZ at $0<H<H_s$ even in the harmonic approximation. This effect stems from the fact that there are two equivalent sublattices whose magnetizations are closer to each other than to the third sublattice. Then, the direction is selected along the bond in the unit cell connecting spins from these two sublattices. In all results presented here and below, magnetizations are closer of equivalent sublattices 1 and 2 denoted in Fig.~\ref{BZfig}(a). As a consequence, spectra differ, for instance, at $M$ and $M'$ points. This spectra asymmetry is small in the Y-phase but it becomes more pronounced at larger $H$ (see Figs.~\ref{spec1n17}, \ref{dsfvphase}(g), \ref{dsfvphase}(i), and \ref{dsf44} below). This effect may well be an artifact of the first order in $1/n$ and it may disappear upon taking into account all terms in $1/n$ series. On the other hand, the appearance of such high-energy spectra asymmetry does not contradict anything. Further consideration of this model by other methods is required to clarify this point.

\begin{figure}
\includegraphics[scale=0.7]{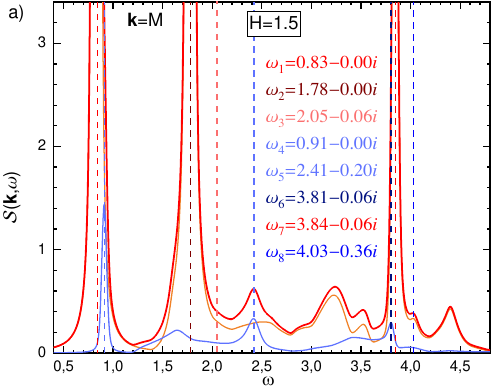}
\includegraphics[scale=0.7]{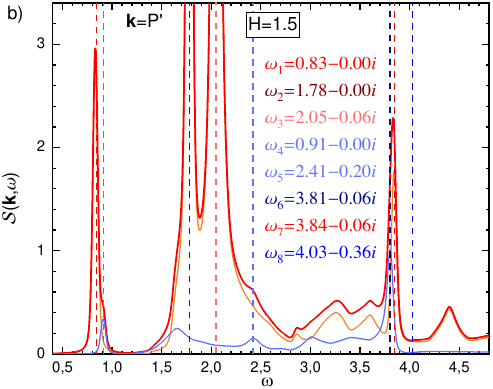}
\includegraphics[scale=0.7]{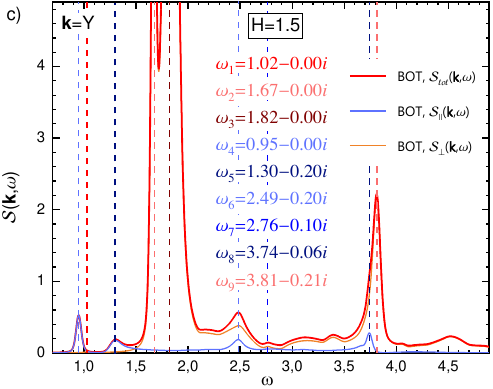}
\caption{
Dynamical structure factors (DSFs) given by Eqs.~\eqref{dsfzz}--\eqref{dsftot} at points $M$, $P'$, and $Y$ of the BZ (see Fig.~\ref{BZfig}(b)) obtained within the first order in $1/n$ in the UUD phase at $H=1.5$ and convoluted with the energy resolution of $0.02J$. Anomalies in DSFs are produced by poles of the respective spin correlators indicated in insets by colors corresponding to excitation branches shown in Figs.~\ref{spec01} and \ref{spec1n17}. Real parts of these poles are marked by vertical dashed lines of respective colors. It is seen that there are poles stemming from the same pole in the harmonic approximation of the BOT.
\label{dsf15}}
\end{figure}

\underline{$H_1<H<H_2$.} It is natural that transverse \eqref{dsfperp} and longitudinal \eqref{dsfzz} DSFs differ qualitative in the collinear UUD phase. It is seen from Fig.~\ref{dsf15} drawn for $M$, $P'$, and $Y$ points at $H=1.5$ that three magnons produce peaks only in ${\cal S}_\perp({\bf k},\omega)$. Anomalies in ${\cal S}_\|({\bf k},\omega)$ are produced by poles some of which stem from the same poles in the harmonic approximation. Notice very small spectral weights of anomalies in the longitudinal DSF compared to spectral weights produced by magnons in the transverse DSF. Then, it would be practically impossible to study the spin-0 excitations in experiments with non-polarized neutrons. 

Remarkably, the lowest-energy spin-0 quasiparticle is well-defined and its spectrum is very close to the spectrum of the lowest magnon. Deep in the UUD phase, this spin-0 elementary excitation lies even below all magnon branches as it is seen from Fig.~\ref{spec1n17} in which we plot spectra of all quasiparticles at $H=1.7$. It is clear from Fig.~\ref{spec1n17} that the remaining three spin-0 excitations show noticeable damping so that they can be considered to be well-defined only in a limited part of the BZ.

\begin{figure}
\includegraphics[scale=1.04]{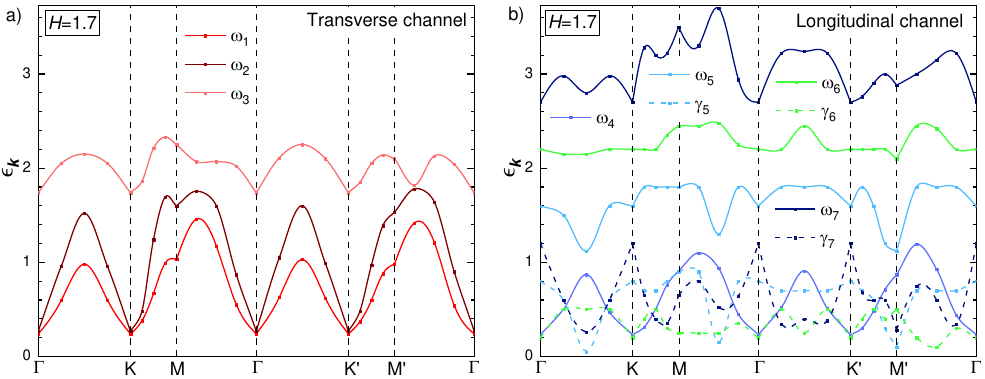}
\caption{
Spectra of low-lying elementary excitations found in the first order in $1/n$ in the UUD phase at $H=1.7$ as it is explained in the text. (a) Spin-1 quasiparticles in the transverse channel which are described by poles $\omega_{1,2,3}$ of transverse dynamical susceptibility \eqref{chiperp}. These elementary excitations are long-lived low-energy magnons corresponding to those in Fig.~\ref{spec01} and drawn in the same colors. (b) Spin-0 quasiparticles of the longitudinal channel. Elementary excitations corresponding to poles $\omega_{4,5,6}$ stem from the mode in the harmonic approximation which is depicted in Fig.~\ref{spec01} by the same color as $\omega_4$. Mode $\omega_7$ originates from the mode shown in Fig.~\ref{spec01} by the same color. Quasiparticles $\omega_{5,6,7}$ have finite damping $\gamma_{5,6,7}$ shown by dashed lines of corresponding colors.
\label{spec1n17}}
\end{figure}

The spectra asymmetry discussed above is clearly seen in Fig.~\ref{spec1n17} (compare spectra, e.g., at points $M$ and $M'$).

\underline{$H_2<H<H_s$.} Apart from the nontrivial evolution of branches appearing even in the harmonic approximation and discussed above, we observe new short-wavelength quasiparticles (new poles in spin correlators) in the first order in $1/n$ in the V phase. Fig.~\ref{dsfvphase} illustrates these our findings in which DSFs are shown at points $M$, $P'$, and $Y$ of BZ (see Fig.~\ref{BZfig}(b)) at $H=3.0$, 4.0, and 4.4. Anomalies in DSFs are produced by poles of the respective spin correlators indicated in insets by colors corresponding to excitation branches shown in Figs.~\ref{spec01} and \ref{spec02}. Poles of the same color stem from the same pole in the harmonic approximation of the BOT. There are also poles shown in Fig.~\ref{dsfvphase} in magenta (e.g., pole $\omega_2$ in Fig.~\ref{dsfvphase}(e)) which cannot be related with any pole in the harmonic approximation. 

\begin{figure}
\includegraphics[scale=0.7]{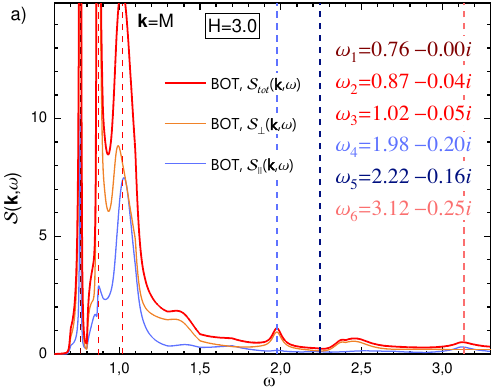}
\includegraphics[scale=0.7]{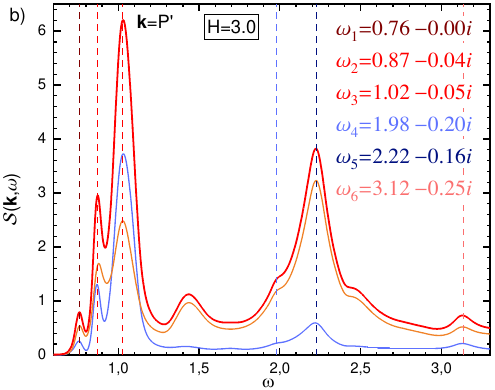}
\includegraphics[scale=0.7]{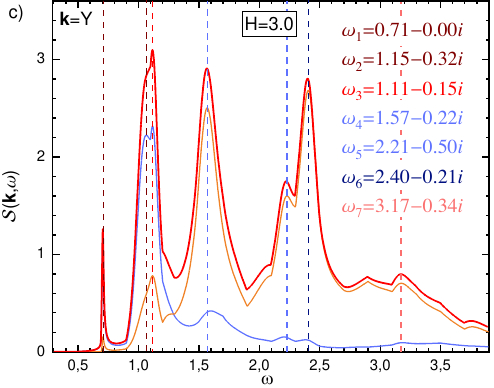}
\includegraphics[scale=0.7]{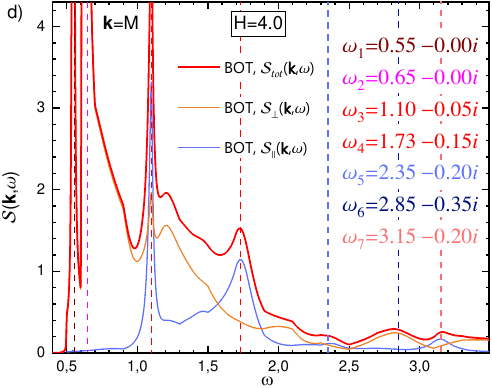}
\includegraphics[scale=0.7]{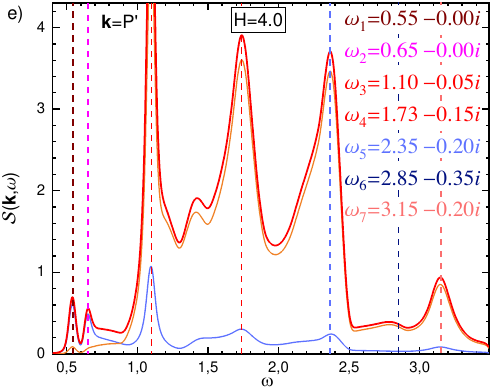}
\includegraphics[scale=0.7]{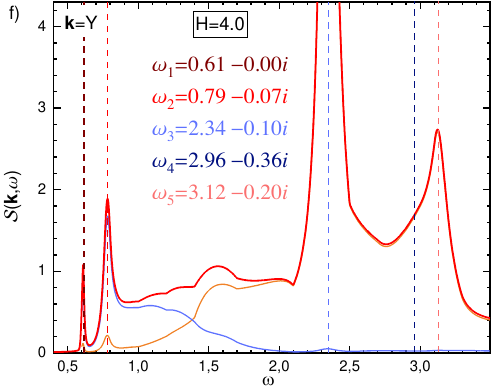}
\includegraphics[scale=0.7]{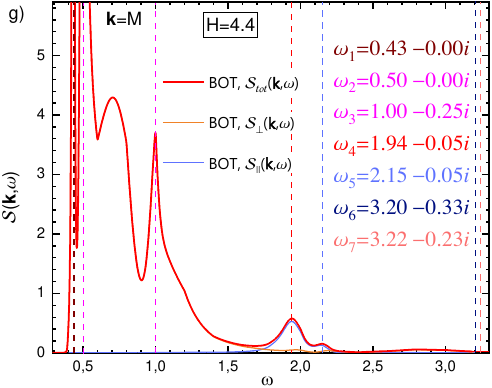}
\includegraphics[scale=0.7]{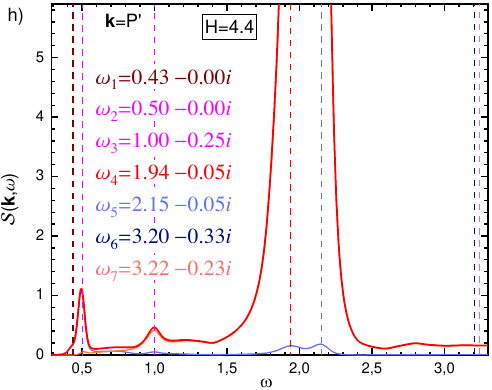}
\includegraphics[scale=0.7]{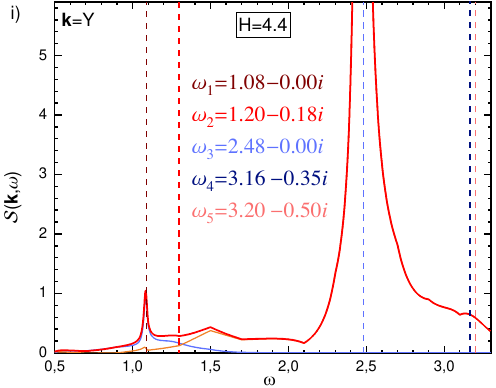}
\caption{
Dynamical structure factors (DSFs) given by Eqs.~\eqref{dsfzz}--\eqref{dsftot} at points $M$, $P'$, and $Y$ of the BZ (see Fig.~\ref{BZfig}(b)), obtained within the first order in $1/n$ at $H=3.0$, 4.0, and 4.4 (i.e., in phase V). Anomalies in DSFs are produced by poles of the respective spin correlators indicated in insets by colors corresponding to excitation branches shown in Figs.~\ref{spec01} and \ref{spec02}. Real parts of these poles are marked by vertical dashed lines of respective colors. Poles of the same color stem from the same pole in the harmonic approximation of the BOT. Poles shown in magenta (e.g., pole $\omega_2$ in panel (d)) cannot be related with any pole in the harmonic approximation.
\label{dsfvphase}}
\end{figure}

The absence of the sixfold rotation symmetry of spectra which is mentioned above can be noted even near the saturation field by comparing Figs.~\ref{dsfvphase}(g) and \ref{dsfvphase}(i) with Fig.~\ref{dsf44} plotted at $H=4.4$ for $M$, $Y$ and $M'$, $Y'$ points, respectively.

\begin{figure}
\includegraphics[scale=1.0]{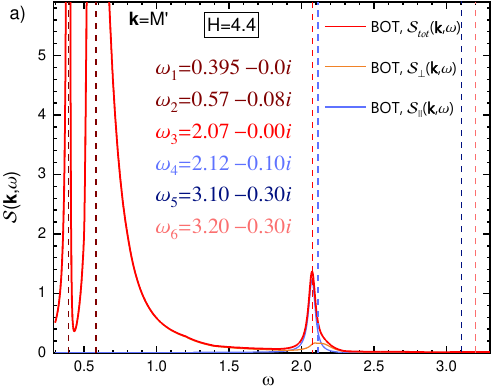}
\includegraphics[scale=1.0]{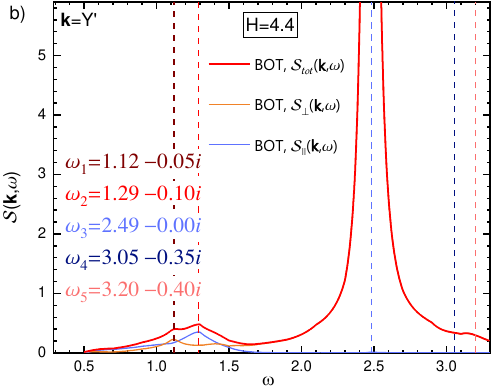}
\caption{Same as Fig.~\ref{dsfvphase} but for $M'$ and $Y'$ points at $H=4.4$.
\label{dsf44}}
\end{figure}

\begin{figure}
\includegraphics[scale=1.1]{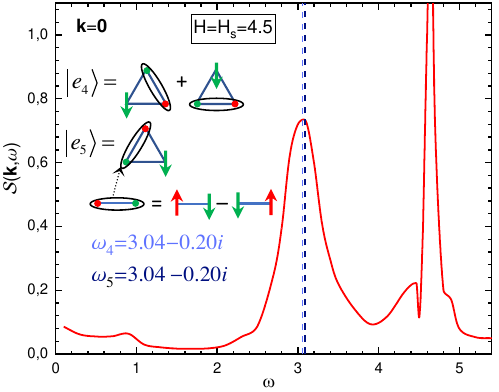}
\caption{
Dynamical structure factor (DSF) ${\cal S}({\bf k=0},\omega)=\frac1\pi{\rm Im}\chi({\bf k=0},\omega)$ at $H=H_s$ found in the first order in $1/n$, where $\chi({\bf k},\omega)$ is the four-spin susceptibility \eqref{chiex} built on spin operators \eqref{operex}. These operators are given in the leading order in $1/n$ by bosonic operators creating spin-2 excitations $|e_4\rangle$ and $|e_5\rangle$ shown in the inset. These two bosons produce two degenerate dispersionless branches with energy $\omega=3.5$ in the harmonic approximation (see Fig.~\ref{spec02}(i)). They give the broad peak at $\omega\approx3.1$ in the DSF in the first order in $1/n$ corresponding to poles $\omega_{4,5}$ of the DSFs. These spin-2 excitations appear in the SWT as bound states of two magnons. The narrow peak in the DSF at $\omega\approx 4.6$ does not correspond to a pole of $\chi({\bf k},\omega)$. 
\label{spechs}}
\end{figure}

\underline{$H=H_s$.} It is seen from Fig.~\ref{spec02}(i) that there are two degenerate dispersionless branches of spin-2 excitations with energy $\omega=3.5$ at $H=H_s$. While these modes lie below the magnon branch around $\Gamma$ point, their spectral weights are zero at $H\ge H_s$ in two-spin correlators \eqref{chi} (see Fig.~\ref{spec02}(j)). However, these excitations become apparent at $H\ge H_s$ in the four-spin correlators of the type
\begin{equation}
\label{chiex}
\chi({\bf k},\omega) =
i\int_0^\infty dt 
e^{i\omega t}	
\left\langle \left[ {\cal A}_{\bf k}(t), {\cal A}^\dagger_{-\bf k}(0) \right] \right\rangle,
\end{equation}
where ${\cal A}_{\bf k}$ are linear combinations of products of two operators $S^-$. For instance, let us consider operators
\begin{eqnarray}
\label{operex}
{\cal A}_{1j}^\dagger &=& \frac{1}{\sqrt6} \left( 2S_{1j}^-S_{2j}^- - S_{1j}^-S_{3j}^- - S_{2j}^-S_{3j}^- \right ),
\nonumber\\
{\cal A}_{2j}^\dagger &=& \frac{1}{\sqrt2} \left( S_{1j}^-S_{3j}^- - S_{2j}^-S_{3j}^- \right ),
\end{eqnarray}
where ${\bf S}_{pj}$ is the $p$-th spin in the $j$-th unit cell (see Fig.~\ref{BZfig}(a)). It can be shown that bosonic representations of ${\cal A}_{1j}^\dagger$ and ${\cal A}_{2j}^\dagger$ constructed as it is described in Ref.~\cite{ibot} are equal in the leading order in $1/n$ to bosonic operators creating the considered excitations. Then, spin susceptibility \eqref{chiex} is related to the Green's functions of the considered bosons. Our calculations of these two Green's functions show that they remain equal to each other in the first order in $1/n$ and the corresponding DSF at $\bf k=0$ is shown in Fig.~\ref{spechs}. It is seen that these excitations produce distinct anomaly in the four-spin correlators at $\omega\approx3.1$ and they acquire finite damping in the first order in $1/n$ due to the decay into two magnons. Similar spin-2 boson lying below the magnon band around $\Gamma$ point was found also at $H\approx H_s$ in the spin-$\frac12$ HAF on the square lattice. \cite{iboth} The narrow peak in Fig.~\ref{spechs} at $\omega\approx 4.6$ does not correspond neither to a pole of $\chi({\bf k},\omega)$ nor to an edge of two-magnon continuum. It appears due to a reduction of the denominator of $\chi({\bf k},\omega)$ which, however, does not vanish upon varying imaginary $\omega$ at $|\omega|\sim4.6$ (in contrast, the denominator vanishes at $\omega\approx3.04-0.20i$).

\section{Comparison with experiment}
\label{expersec}

\subsection{$\rm Ba_3CoSb_2O_9$}

In this section, we take into account a small easy-plane anisotropy $A>0$ and consider the system on the stacked triangular lattice having the Hamiltonian (cf.\ Eq.~\eqref{ham})
\begin{equation}
\label{hamanis}
{\cal H} = \sum_{\langle i,j \rangle}	J\left( {\bf S}_i{\bf S}_j - A S_i^y S_j^y \right) - H\sum_i S_i^z + {\cal H}_{3D},
\end{equation} 
where ${\cal H}_{3D}$ stands for a small exchange coupling between spins from nearest triangular planes with exchange constant $J'$.
As it was established before, \cite{bacoprl,bacogap,bacoH} model \eqref{hamanis} describes well $\rm Ba_3CoSb_2O_9$ with $J\approx1.7$~meV, $A\approx0.1$, $J'\approx0.05J$, and nearly isotropic $g$-factor of 3.85. In our previous study \cite{itri} of model \eqref{hamanis} at $H=0$, we neglect ${\cal H}_{3D}$ for simplicity and find a good quantitative agreement with many experimental observations at
\begin{equation}
\label{param}
J=1.77\,{\rm meV}, \quad A=0.15.
\end{equation}
Magnon spectra were described quantitatively in the UUD phase in Ref.~\cite{bacoH} within the first order in $1/S$ using model \eqref{hamanis} with parameters \eqref{param} and $J'\approx0.09J$. To describe available neutron data reported in Ref.~\cite{bacoH} in the UUD phase, we use the same set of parameters \eqref{param} and calculate the following dynamical structure factor: \cite{Lowesey}
\begin{equation}
\label{neutron}
{\cal S}({\bf k},\omega) = 
\frac1\pi  {\rm Im}
\sum_{\alpha,\beta} 
g_{\alpha}g_{\beta}
\left(
\delta_{\alpha\beta} - \widehat k_\alpha \widehat k_\beta
\right)
\chi_{\alpha\beta}({\bf k},\omega),
\end{equation}
where $\alpha,\beta=x,y,z$, $\widehat{\bf k}={\bf k}/k$, $g_{\alpha}$ is the $g$-tensor component, and $\chi_{\alpha\beta}({\bf k},\omega)$ are given by Eq.~\eqref{chi}. 

\begin{figure}
\includegraphics[scale=0.7]{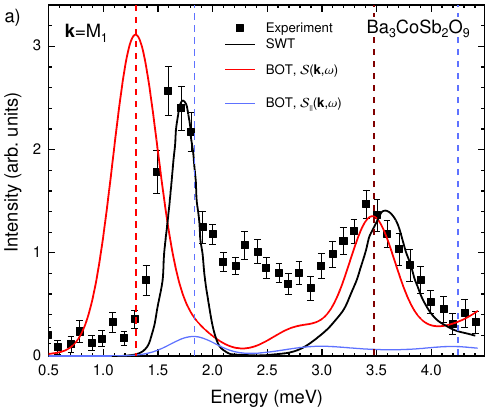}
\includegraphics[scale=0.7]{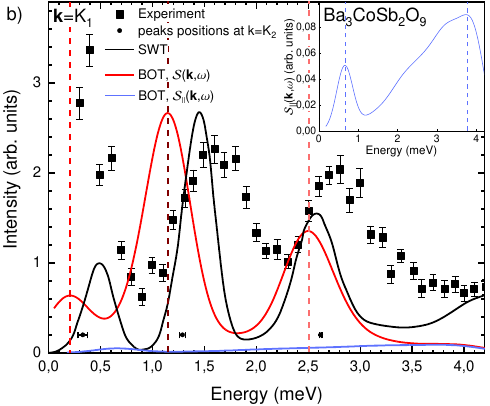}
\includegraphics[scale=0.71]{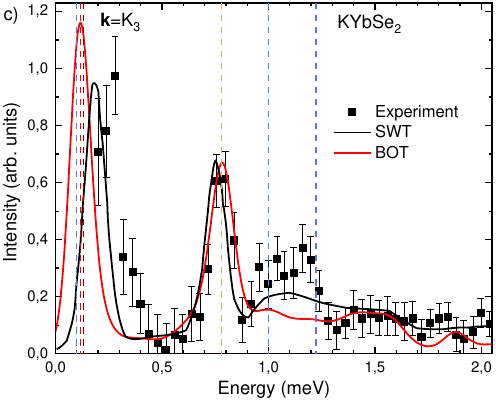}
\caption{
(a) and (b) Inelastic neutron scattering intensity obtained experimentally \cite{bacoH} at $H=10.5$~T in the UUD phase of $\rm Ba_3CoSb_2O_9$ at $M_1$ and $K_1$ points in the BZ corresponding to ${\bf k}=(3/2,3/2,-2)$ and ${\bf k}=(1,1,-2)$, respectively. Positions of experimentally observed \cite{bacoH} peaks at $K_2$ point (${\bf k}=(1,1,1)$) are also shown in panel (b) (see discussion in the text). Results are presented of spin-wave calculations from Ref.~\cite{bacoH} (SWT). Red curves are theoretical results of the present study with parameters \eqref{param} in model \eqref{hamanis} convoluted with the experimental energy resolution 0.2~meV. Vertical dashed lines indicate real parts of poles of spin correlator as in Fig.~\ref{dsf15}. Inset in panel (b) shows contribution to DSF \eqref{neutron} from $\chi_{zz}({\bf k},\omega)$ whose anomalies correspond to novel quasiparticles (see also Fig.~\ref{spec1n17}). (c) Same as (a) and (b) but for $K_3$ point corresponding to ${\bf k}=(1,1,3)$ in $\rm KYbSe_2$ at $H=4$~T ($H\approx1.7J$). Experimental data and SWT results in panel (c) are taken from Fig.~S5 of Ref.~\cite{kybse}.
\label{experiment}}
\end{figure}

Results of our calculations for $M_1$ (${\bf k}=(3/2,3/2,-2)$) and $K_1$ (${\bf k}=(1,1,-2)$) points at $H=10.5$~T (i.e., in the UUD state) are shown in Figs.~\ref{experiment}(a) and \ref{experiment}(b) together with neutron data and results of previous spin-wave calculations from Ref.~\cite{bacoH}. A good agreement is seen between our findings, the SWT and the experiment. Discrepancies between our results and the experiment can be attributed to the neglect of the inter-plane interaction in our calculations. The influence of $J'$ can be estimated as the difference between magnon peaks positions at $K_1$ and $K_2$ (${\bf k}=(1,1,1)$) points which differ in the last coordinate. Because $J'$ is antiferromagnetic, the periodicity of the spectrum in the direction perpendicular to the triangular planes is twice as large as the periodicity of the reciprocal lattice. Positions of experimentally observed \cite{bacoH} peaks at $K_2$ are depicted in Fig.~\ref{experiment}(b). It is seen that small discrepancies between our findings, results of the SWT and experimental data can be well ascribed to the neglect of the inter-plane interaction. 

The contribution to DSF \eqref{neutron} from $\chi_{zz}({\bf k},\omega)$ is also presented in Figs.~\ref{experiment}(a) and \ref{experiment}(b) and it is seen that the longitudinal DSF cannot be extracted from these experimental data due to its minor contribution.

\subsection{$\rm KYbSe_2$}

$\rm KYbSe_2$ is believed \cite{kybse} to be described by model \eqref{ham} with a small extra exchange coupling $J_2$ between next-nearest spins and negligible inter-layer interaction. Then, the model to be discussed has the form
\begin{equation}
\label{hamj2}
{\cal H} = J\sum_{\langle i,j \rangle}{\bf S}_i{\bf S}_j 
+ 
J_2\sum_{\langle \langle i,j \rangle \rangle}{\bf S}_i{\bf S}_j 
- 
H\sum_i S_i^z.
\end{equation} 
It was proposed in Ref.~\cite{kybse} that $J\approx0.46$~meV and $J_2\approx0.043J$. We assume that $g$-tensor is diagonal with components $g_{ab}=3.41$ and $g_c=0.65$ within the triangular planes and perpendicular to them, respectively. \cite{kybse,kybse2}

Results of our calculations for $K_3$ point (${\bf k}=(1,1,3)$) at $H=4$~T (corresponding to $H\approx1.7J$ in the UUD phase) are shown in Fig.~\ref{experiment}(c), where a good agreement is seen of our findings with neutron data and results of spin-wave calculations from Ref.~\cite{kybse}. We restrict ourselves to this comparison in $\rm KYbSe_2$ and turn to a more detailed comparison with neutron data in the isostructural material $\rm CsYbSe_2$ reported in Ref.~\cite{csybse2}, where authors managed to separate longitudinal and transverse DSFs using polarized neutrons.

\subsection{$\rm CsYbSe_2$}

Numerical investigations performed in Refs.~\cite{csybse1,csybse2} show that model \eqref{hamj2} describes well neutron scattering data in $\rm CsYbSe_2$ at $J\approx0.4$~meV, $J_2\approx0.03J$, and with $g$-tensor components $g_{ab}=3.25$ and $g_c=0.3$.

\begin{figure}
\includegraphics[scale=0.85]{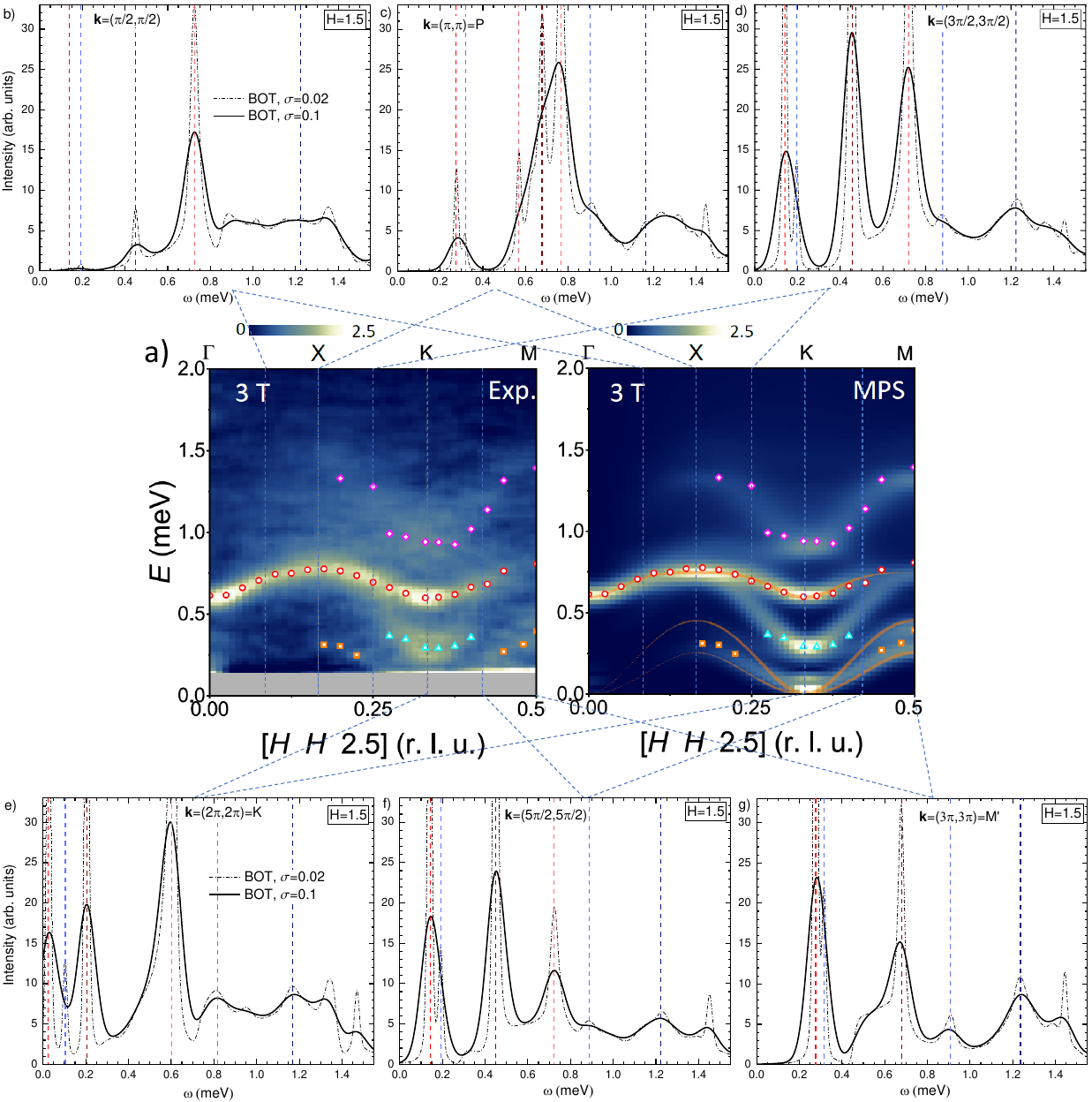}
\caption{(a) Neutron scattering data (Exp.) and numerical findings obtained using matrix-product-state (MPS) representations reported in Ref.~\cite{csybse2} for $\rm CsYbSe_2$ at $H=3$~T in the UUD state (all density plots are taken from Fig.~3 in Ref.~\cite{csybse2}). (b)--(g) Results of calculation within the BOT of DSF \eqref{neutron} in model \eqref{hamj2} at six representative momenta at $H=1.5J$ (corresponding to $H=3$~T). Vertical dashed lines indicate real parts of poles of spin correlators as in Fig.~\ref{dsf15}. There are two curves in each panel showing BOT results convoluted with two values of the energy resolution $\sigma$: $0.02J$ and $0.1J$. The latter value is close to the experimental and numerical resolution in Ref.~\cite{csybse2}.
\label{yb3t}}
\end{figure}

\begin{figure}
\includegraphics[scale=0.85]{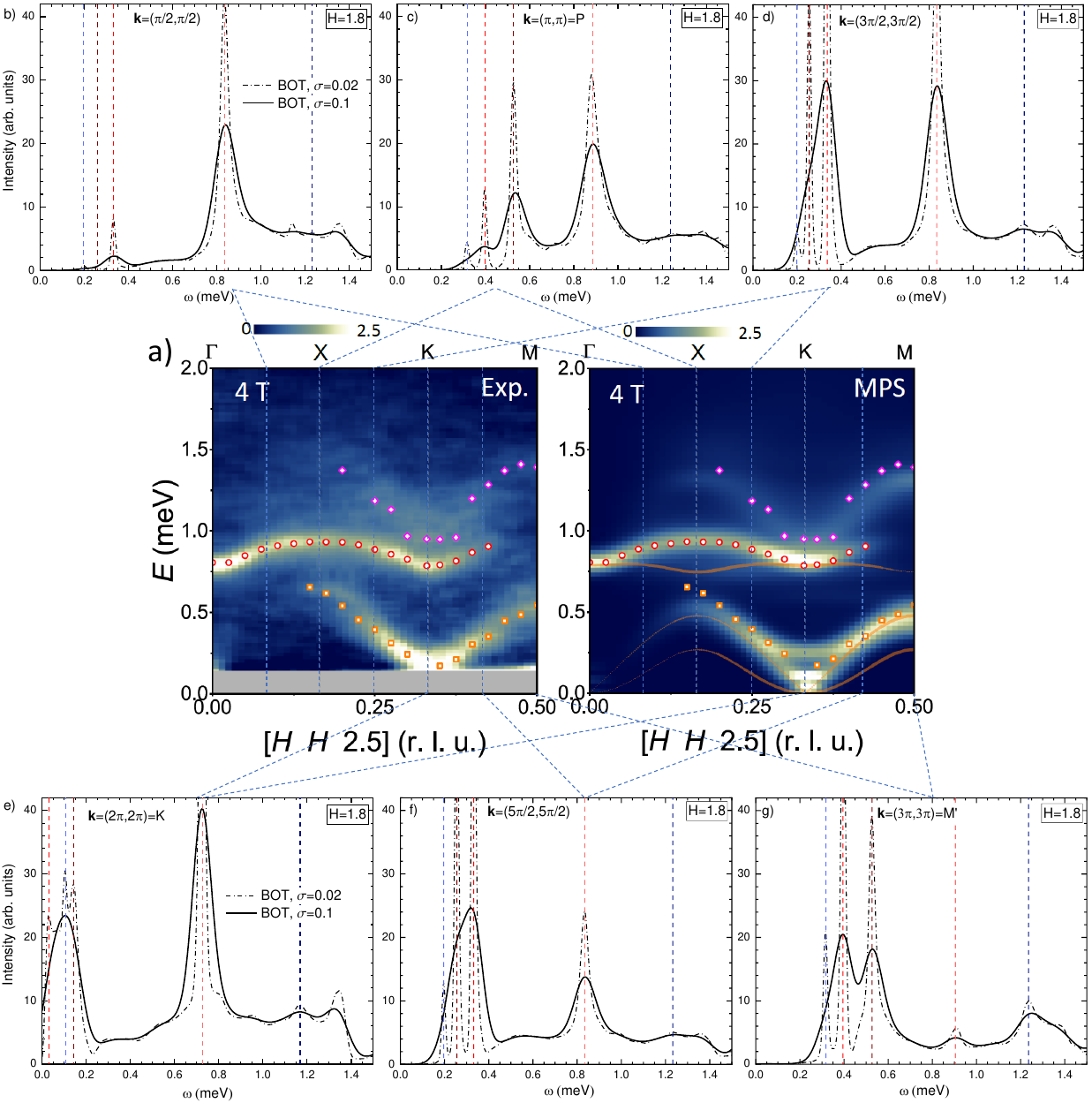}
\caption{Same as Fig.~\ref{yb3t} but for $H=4$~T (corresponding to $H=1.8J$). All density plots in panel (a) are taken from Fig.~3 in Ref.~\cite{csybse2}.
\label{yb4tall}}
\end{figure}

\begin{figure}
\includegraphics[scale=0.85]{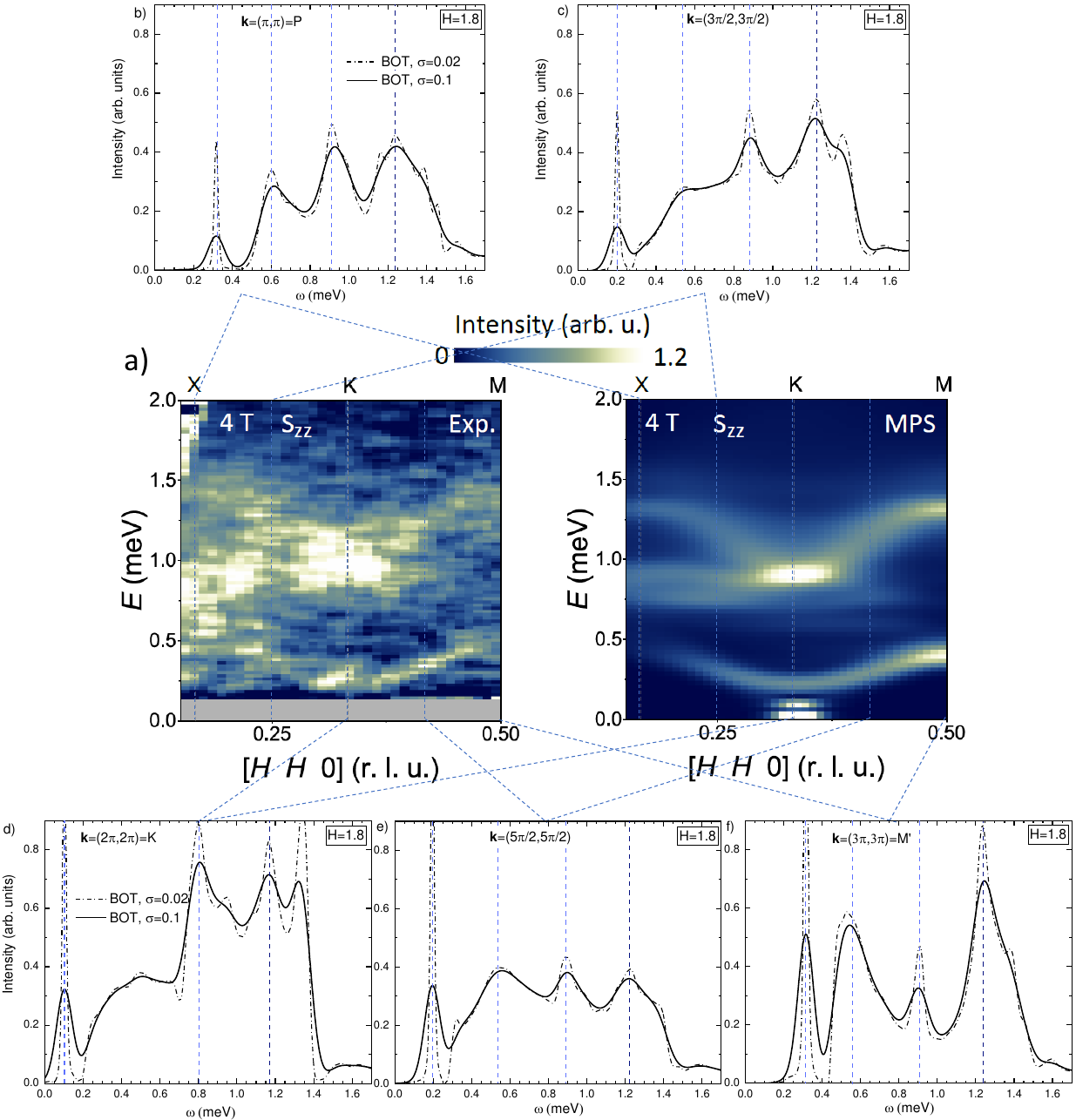}
\caption{(a) Numerical findings obtained using matrix-product-state (MPS) representations and neutron scattering data for the longitudinal correlator reported in Ref.~\cite{csybse2} for $\rm CsYbSe_2$ at $H=4$~T in the UUD state (all density plots are taken from Fig.~4 in Ref.~\cite{csybse2}). (b)--(f) Same as in Figs.~\ref{yb3t}(b)--(g) but for DSF \eqref{dsfzz} in model \eqref{hamj2} at $H=1.8J$ (corresponding to $H=4$~T). 
\label{yb4tlong}}
\end{figure}

We present results of our calculation of DSF \eqref{neutron} for $\rm CsYbSe_2$ at $H=3$~T and 4~T (the UUD phase) in Figs.~\ref{yb3t} and \ref{yb4tall}, respectively. It is seen from these figures that BOT reproduces quite accurately all characteristic spectral features observed both experimentally and numerically in Ref.~\cite{csybse2}. The worse (but still reasonably good) agreement is in positions and intensities of anomalies originating from spin-0 excitations. This is best seen in Fig.~\ref{yb4tlong} demonstrating longitudinal DSFs. However we can conclude that the number of main anomalies (corresponding to the number of spin-0 quasiparticles found using the BOT), the anomalies positions (corresponding to energies of spin-0 quasiparticles), and their relative intensities are in agreement with numerical findings.

\begin{figure}
\includegraphics[scale=0.85]{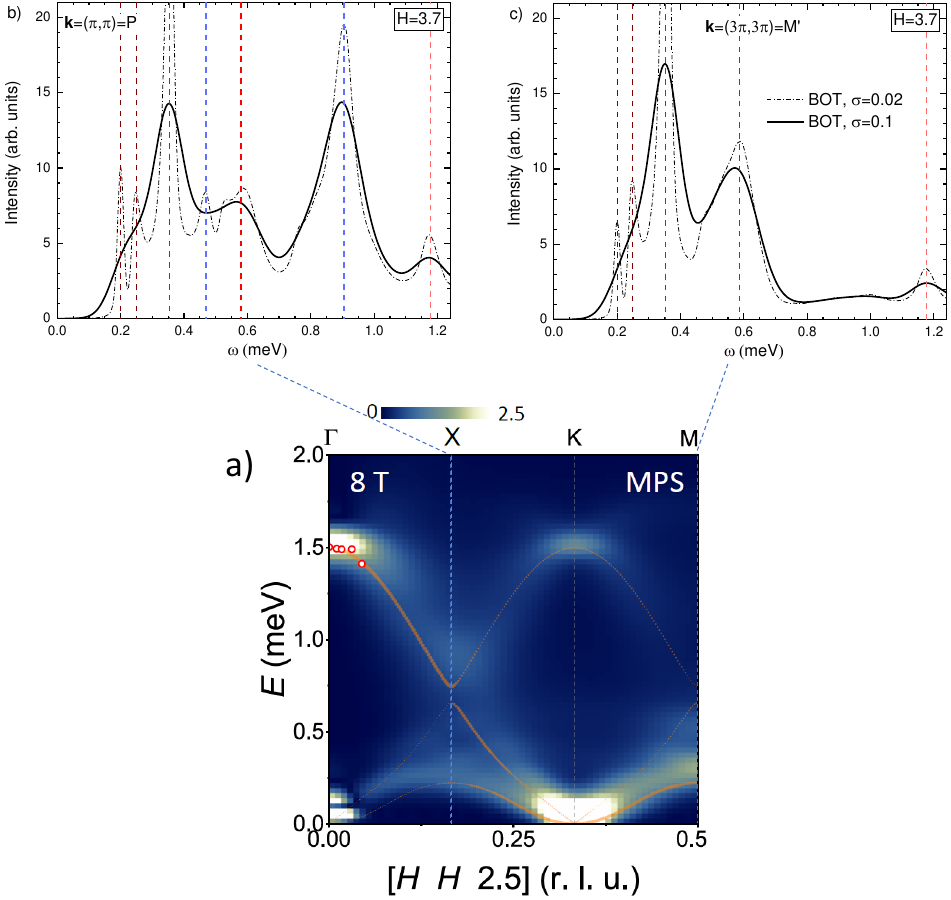}
\caption{Same as Fig.~\ref{yb3t} but for $H=8$~T (corresponding to $H=3.7J$) in the V-phase. The density plot in panel (a) is taken from Fig.~3 in Ref.~\cite{csybse2}. The appearance in panels (b) and (c) of the anomaly at $\omega\approx0.58$~meV produced by new quasiparticle is quite consistent with numerical results shown in slide (a).
\label{yb8t}}
\end{figure}

Of particular interest is also the V-phase in which novel quasiparticles produce sharp anomalies standing separately (see, e.g., Fig.~\ref{dsfvphase}(e) for $P'$ point). We show in Fig.~\ref{yb8t} that the appearance in the BOT calculations of the anomaly at $\omega\approx0.58$~meV produced by new quasiparticle is quite consistent with numerical results at $H=8$~T (corresponding to $H=3.7J$).

\section{Summary and conclusion}
\label{conc}

To conclude, we discuss dynamical properties of spin-$\frac12$ Heisenberg antiferromagnet on the triangular lattice in magnetic field \eqref{ham}. We use the bond-operator theory (BOT) which accurately takes into account short-range spin correlations and provides a quantitative description of excitations which appear in other approaches as bound states of conventional magnons. Observable quantities can be found within the suggested variant of the BOT as series in $1/n$, where $n$ is the maximum number of bosons which can occupy a unit cell. Although the physical results correspond to $n=1$, we find in the present study as well as in our previous applications of this approach to other models \cite{ibot,aktersky,iboth,itri} that first terms in $1/n$ provide the main corrections to most observables bringing our results to quantitative agreement with available previous numerical, experimental, and analytical findings. 

In quantitative agreement with previous results, we observe four phases in this model which are depicted in Fig.~\ref{BZfig}(c). In particular, the collinear UUD state showing the magnetization plateau at 1/3 of its saturation value is stable between critical fields given by Eq.~\eqref{h12} in the first order in $1/n$. As it is seen from Fig.~\ref{emfig}, the ground state energy and the uniform magnetization obtained within the BOT in the first order in $1/n$ are in good quantitative agreement with previous numerical findings.

We obtain that at small field (in phase Y) three low-energy branches of excitations correspond to conventional magnons known from the spin-wave theory (SWT). However, we demonstrate using the BOT in Ref.~\cite{itri} that quantum fluctuations lift magnon spectra degeneracy at $H=0$ along $\Gamma M$ lines and along dashed lines shown in Fig.~\ref{BZfig}(b) which is predicted by the SWT even in the first order in $1/S$. This our finding is in a quantitative agreement with experimental results in $\rm Ba_3CoSb_2O_9$. \cite{itri} We find also four high-energy excitations which produce high-energy broad peak at $\omega\approx2.5$ in the dynamical structure factor (DSF) at $M$ point of the Brillouin zone (BZ) at $H=0$ (see Fig.~\ref{dsfm005}). As it is also seen from Fig.~\ref{dsfm005}, this peak loses shape upon the field increasing because four high-energy branches move: two of them go up, the energy of the other does not approximately change, and the fourth goes down (see Fig.~\ref{spec01}). The later branch which is shown in light blue in Fig.~\ref{spec01} describes in the collinear UUD phase the low-energy mode carrying spin 0 and appearing only in longitudinal spin correlator \eqref{chizz} (see Fig.~\ref{spec01}(g)). The pole of spin correlator \eqref{chi} corresponding to this mode in the harmonic approximation of the BOT turns into three poles upon taking into account corrections to the self-energy parts of the first order in $1/n$ and upon reducing $n$ from infinity (harmonic approximation shown in Fig.~\ref{spec01}(g)) to $n=1$ (see Fig.~\ref{spec1n17}). One of these spin-0 quasiparticles is well-defined and it lies below magnon branches deep in the UUD state. The remaining spin-0 modes obey quite a strong damping in the major part of the BZ (see Figs.~\ref{dsf15} and \ref{spec1n17}). All spin-0 modes produce anomalies in the longitudinal spin correlator whose spectral weights, however, are much smaller than spectral weights of magnons in the transverse spin correlator that hinders its experimental observation (see Figs.~\ref{dsf15} and \ref{experiment}). Notice that these spin-0 excitations could be obtained in the SWT as two-magnon bound states.

Upon further field increasing (in phase V), we find a nontrivial spectra modification which can be demonstrated even in the harmonic approximation of the BOT using Figs.~\ref{spec01} and \ref{spec02}. It is seen that the branch, which is shown in light blue in Figs.~\ref{spec01} and \ref{spec02} and which describes purely spin-0 excitations in the UUD phase, becomes a hybrid at $H\to H_s$: one its part becomes a part of the magnon spectrum whereas its horizontal part describes spin-2 quasiparticles. Similar modification occurs with the branch shown in orange in Figs.~\ref{spec01} and \ref{spec02} which describes the high-energy magnon in the UUD phase and which turns out to be hybrid at $H\to H_s$. This spectra evolution would be difficult to predict by any conventional approach. Then, in terms of the SWT, we obtain a nontrivial interplay in phase V between one-particle and two-particle sectors which cannot be reproduced in first few orders in $1/S$.

Similar to spin-$\frac12$ HAF on the square lattice in strong magnetic field, we find multiple short-wavelength spin excitations in strong field (in phase V) some of which have no counterparts neither in the SWT nor in the harmonic approximation of the BOT (see Fig.~\ref{dsfvphase}). 

We observe also the lack of the sixfold rotation symmetry of high-energy parts of spectra at $0<H<H_s$ which does not follow from the linear SWT. It is related with the fact that there are two equivalent sublattices whose staggered magnetizations are closer to each other than to the magnetization of the third sublattice. As a consequence, the direction is selected in the unit cell connected spins with closer magnetizations that is reflected in excited states of the unit cell considered in the BOT. This spectra asymmetry can be seen in Figs.~\ref{spec01}, \ref{spec02}, \ref{spec1n17}, \ref{dsfvphase}(g), \ref{dsfvphase}(i), and \ref{dsf44} (magnetizations of equivalent sublattices 1 and 2 denoted in Fig.~\ref{BZfig}(a) are closer in presented results). Further consideration of this model by other methods is required to clarify whether this high-energy spectra asymmetry is an artifact of the first order in $1/n$ or it is a physical effect.

Our main findings are in good agreement with available experimental and numerical data observed in $\rm Ba_3CoSb_2O_9$ (Figs.~\ref{experiment}(a) and \ref{experiment}(b)), $\rm KYbSe_2$ (Fig.~\ref{experiment}(c)), and $\rm CsYbSe_2$ (Figs.~\ref{yb3t}, \ref{yb4tall}, \ref{yb4tlong}, and \ref{yb8t}). In particular, the appearance in the BOT of multiple spin-0 quasiparticles in the UUD state and novel quasiparticles in the V-phase are consistent with recent experimental and numerical results presented in Figs.~\ref{yb4tlong} and \ref{yb8t}.

\begin{acknowledgments}

This work is supported by the Russian Science Foundation (Grant No.\ 22-22-00028). 

\end{acknowledgments}

\bibliography{tribib}

\end{document}